\documentclass[12pt]{article}
\pdfoutput=1
\usepackage{graphicx}
\usepackage{epsfig}
\usepackage{epstopdf}
\DeclareGraphicsExtensions{.pdf,.eps,.png,.jpg,.mps}
\usepackage{caption}
\usepackage{float}
\usepackage{color}

\usepackage{url}
\usepackage[implicit=false]{hyperref}
\usepackage{cite}

\def\lsim{\mathrel{\rlap{\lower3pt\hbox{\hskip0pt$\sim$}}
     \raise1pt\hbox{$<$}}}         %less than or approx. symbol
\def\gsim{\mathrel{\rlap{\lower4pt\hbox{\hskip1pt$\sim$}}
     \raise1pt\hbox{$>$}}}         %greater than or approx. symbol

\usepackage{amsmath}
\usepackage{amsfonts}

\begin{document}
\begin{titlepage}

\centerline{\Large \bf Healthy\dots Distress\dots Default }
\medskip

\centerline{Zura Kakushadze$^\S$$^\dag$\footnote{\, Zura Kakushadze, Ph.D., is the President of Quantigic$^\circledR$ Solutions LLC,
and a Full Professor at Free University of Tbilisi. Email: \href{mailto:zura@quantigic.com}{zura@quantigic.com}}}
\bigskip

\centerline{\em $^\S$ Quantigic$^\circledR$ Solutions LLC}
\centerline{\em 1127 High Ridge Road \#135, Stamford, CT 06905\,\,\footnote{\, DISCLAIMER: This address is used by the corresponding author for no
purpose other than to indicate his professional affiliation as is customary in
publications. In particular, the contents of this paper
are not intended as an investment, legal, tax or any other such advice,
and in no way represent views of Quantigic$^\circledR$ Solutions LLC,
the website \url{www.quantigic.com} or any of their other affiliates.
}}
\centerline{\em $^\dag$ Free University of Tbilisi, Business School \& School of Physics}
\centerline{\em 240, David Agmashenebeli Alley, Tbilisi, 0159, Georgia}
\medskip
\centerline{(June 4, 2019)}

\bigskip
\medskip

\begin{abstract}
{}We discuss a simple, exactly solvable model of stochastic stock dynamics that incorporates regime switching between healthy and distressed regimes. Using this model, which is analytically tractable, we discuss a way of extracting expected returns for stocks from realized CDS spreads, essentially, the CDS market sentiment about future stock returns. This alpha/signal could be useful in a cross-sectional (statistical arbitrage) context for equities trading.
\end{abstract}
\medskip
\bigskip
\bigskip
\bigskip
{}{\bf Keywords:} stock, CDS spread, healthy, distress, default, stochastic dynamics, statistical arbitrage, alpha, regime switching, expected return, market sentiment, equities trading

\bigskip
\bigskip
\bigskip

\end{titlepage}

\newpage

{}A geometric Brownian motion (with a drift) is a simple model of stock dynamics used, e.g., in the Black-Scholes-Merton model \cite{BS}, \cite{Merton1973} in the context of options (and other derivatives) pricing. Such a model describes a stock price that on average increases (or decreases) exponentially in time, but does not incorporate any regime-switching between the healthy (increasing stock price) and distressed (decreasing stock price) regimes, with the distressed regime potentially leading to an eventual default. Merton's corporate default model \cite{Merton1974} incorporates an ad hoc threshold for the underlying (unobservable) ``firm value" stochastic process below which a credit event is interpreted to have occurred.

{}In this note we discuss a simple model of stock dynamics that smoothly interpolates between the healthy and distressed regimes. This model is exactly solvable in the sense that, even though the underlying dynamics of the (unobservable) state variable is nontrivial (the drift term for the state variable is a nonlinear function of said variable), the transition density is expressed through elementary functions. A long-run probability of a healthy-to-distressed transition (which can be interpreted as the probability of default) can thus be readily computed. A connection of this model to physics, among other things, is that the Schr{\"o}dinger equation for the transition density turns out to have a constant potential, hence exact solvability.

{}This model can be thought of as a simple (toy) model for CDS pricing (with only 2 free parameters). Conversely, it can be thought of as a model for extracting the CDS market sentiment about the expected return of the stock (of the company whose debt the CDS insures), at least in a cross-sectional, statistical sense, i.e., when applied to a broad cross-section of stocks. To be clear, this is a simple {\em illustrative} model and we make no claim regarding its empirical prowess. More complex models with more parameters can be constructed along these lines, and, as always, one must bear in mind that, while more parameters can improve in-sample fits (``calibration"), often they do not necessarily translate into out-of-sample forecasting power.

{}So, here is a simple 2-parameter model. Let the stock price $S_t$ be modeled via
\begin{eqnarray}
 &&S_t = \exp(X_t)\\
 &&dX_t = \mu(X_t)~dt + \sigma~dW_t\label{stoch}
\end{eqnarray}
The drift $\mu(X_t)$ has no explicit dependence on time $t$, and the volatility $\sigma$ is constant; $W_t$ is a Brownian motion (a.k.a. a Wiener process), with null drift and variance $t$.

{}Let $P(x,x_0;t,0)$ be the probability distribution of starting at $X_t = x_0$ at $t=0$ and ending at $X_t = x$ at time $t$. The Fokker-Planck equation reads:
\begin{equation}
 \partial_t P = {\sigma^2\over 2}~\partial_x^2 P - \partial_x\left[\mu~P\right]
\end{equation}
Let us reduce it to the Schr{\"o}dinger equation via
\begin{equation}
 P = \exp\left[{1\over\sigma^2}~\int_{x_0}^x \mu(x^\prime)~dx^\prime\right] {\widetilde P}
\end{equation}
We have
\begin{eqnarray}
 &&\partial_t {\widetilde P} = {\sigma^2\over 2} \left[\partial_x^2 {\widetilde P} - U~{\widetilde P}\right]\\
 &&U = h^2 + \partial_x h\\
 &&h = {\mu\over \sigma^2}
\end{eqnarray}

{}We can model healthy and distressed regimes by having a smooth transition between asymptotically positive (for $x\rightarrow +\infty$) drift and asymptotically negative (for $x\rightarrow -\infty$) drift. A simple, exactly solvable model is given by:
\begin{eqnarray}\label{mod}
 &&\mu(x) = \nu~\sigma^2~\tanh\left[\nu\left(x - x_*\right)\right]\\
 &&U(x) \equiv \nu^2
\end{eqnarray}
So, the ``potential" $U(x)$ in the Schr{\"o}dinger equation is constant; however, the drift $\mu(x)$ smoothly interpolates between positive (healthy) and negative (distressed) values. The probability density (normalized to 1 when integrated over $x$ from $-\infty$ to $+\infty$) then reads:
\begin{equation}
 P(x,x_0;t,0) = {1\over{\sqrt{2\pi t}~\sigma}}~{\cosh\left[\nu\left(x - x_*\right)\right]\over \cosh\left[\nu\left(x_0 - x_*\right)\right]}~
 \exp\left[-{\left(x - x_0\right)^2\over 2\sigma^2 t} - {\sigma^2\nu^2 t\over 2}\right]
\end{equation}
Asymptotically, we have
\begin{equation}
 P(x \rightarrow \pm \infty,x_0;t,0) \sim {1\over{\sqrt{2\pi t}~\sigma}}~\exp\left[-{\left(x - x_0 \mp {\widetilde\mu} t\right)^2\over 2\sigma^2 t}\right]
\end{equation}
where ${\widetilde\mu} = \nu~\sigma^2$, i.e., asymptotically we have probability densities for Brownian motions with constant drifts $+{\widetilde\mu}$ and $-{\widetilde\mu}$ for $x\rightarrow+\infty$ and $x\rightarrow-\infty$, respectively.

{}The price $S_* = \exp(x_*)$ delineates healthy ($S_t > S_*$, i.e., $X_t > x_*$) and distressed ($S_t < S_*$, i.e., $X_t < x_*$) regimes. Using the probability density $P(x,x_0;t,0)$, we can calculate the probability $P(S_T \leq S_* | S_0 > S_*)$ of starting in the healthy regime at time $t = 0$ and ending up in the distressed regime at time $t = T$, and the probability $P(S_T \geq S_* | S_0 < S_*)$ of starting in the distressed regime at time $t = 0$ and ending up in the healthy regime at time $t = T$, which asymptotically (for large $T$) are given by
\begin{eqnarray}\label{h2d}
 &&P(S_T \leq S_* | S_0 > S_*) \rightarrow \left[1 + \left(S_0\over S_*\right)^{2\nu}\right]^{-1}\\
 &&P(S_T \geq S_* | S_0 < S_*) \rightarrow \left[1 + \left(S_*\over S_0\right)^{2\nu}\right]^{-1}\label{d2h}
\end{eqnarray}
The r.h.s. of Eq. (\ref{h2d}) can be interpreted as the probability of default -- this is the probability of starting in the healthy regime at price $S_0 > S_*$ and ending in the distressed regime as $T\rightarrow\infty$, which for all intents and purposes would imply that the company defaults as it does not get out of the distressed state at large $T$. Since the probability of default is smaller than 1 even at large $T$, this implies that the hazard rate is not constant and for a company in the healthy state it decays with time, which is not surprising taking into account the positive drift in $X_t$ for $X_t > x_*$.

{}The above model has two parameters, $\nu$ and $S_*$. More complex functions $\mu(x)$ with similar properties can be considered (at the expense of exact solvability), including those with many more tunable parameters, which may be used to fit empirical data in-sample (albeit this may not translate into out-of-sample predictive power).

{}Eq. (\ref{stoch}) can be viewed as a Langevin equation
\begin{equation}
 {dX_t\over dt} = F(X_t) + \sigma~\eta_t
\end{equation}
Here $\eta_t = dW_t/dt$ is the white noise, $F(x) = -\partial_x V(x)$ is the external force ($F(x) = \mu(x)$), and $V(x)$ is the potential. In the model (\ref{mod}), we have
\begin{equation}
 V(x) = -\sigma^2 \ln\left(\cosh\left[\nu\left(x - x_*\right)\right]\right) + V_*
\end{equation}
where $V_*$ is an immaterial integration constant ($V_* = V(x_*)$). Asymptotically, as $\nu\left|x - x_*\right| \gg 1$, we have $V(x) \sim -{\widetilde \mu} \left|x - x_*\right|$, so the potential is a $\bigwedge$-shaped wedge smoothed out near $x = x_*$ (the cusp of the wedge). Asymptotically, in the healthy regime there is a constant force driving $X_t$ to higher values (positive drift), while in the distressed regime there is a constant force driving $X_t$ to lower values (negative drift). In a sense, akin to \cite{Halperin}, here we have a ``barrier" separating the healthy and distressed regimes, except that there are no minima on either side of the barrier. Asymptotically, the dynamics to the right (left) of the barrier is that of a geometric Brownian motion with a positive (negative) drift.

{}For the sake of simplicity, let us focus on the healthy regime, away from the ``threshold" $S_*$, such that the probability of default is small. Let $P$ be the default probability by time $T$. For maturity $T$, we can relate the CDS spread $Z$ (assuming it is measured in basis points) to $P$ via $Z \approx 10^4\left(1-R\right) P / T$, where $R$ is the recovery rate (typically, $R = 0.4$). This assumes a constant hazard rate, which is not the case in the model above. However, for small default probabilities we can assume a linear relationship between the CDS spread $Z$ and the default probability $P$. The ``tricky" part is the normalization factor as it involves the maturity $T$, whereas in Eq. (\ref{h2d}) we take the large $T$ limit. Happily, as we will see in a moment, for our purposes here the precise normalization factor is actually not needed. Furthermore, the large $T$ limit result (\ref{h2d}) is valid so long as $\sigma\nu\sqrt{T} \gg 1$, that is, ${\widetilde\mu}~T \gg \sigma\sqrt{T}$.

{}So, we can try to think about our model as a model for CDS pricing. While such (or similar) models might work well in-sample (``calibration"), their predictive power out-of-sample typically is at best questionable. This is because at least one parameter -- in this case $\nu$ -- is hard to predict accurately out-of-sample based on historical data. Indeed, predicting $\nu$ is equivalent to predicting the drift, i.e., the expected return of the stock. Put another way, this in fact is equivalent to having a highly predictive alpha model for the stock, which is no easy feat to accomplish.

{}Alternatively, we can turn the tables and think of our model as a means of extracting the stock expected return from the CDS spread $Z$. Assuming that the default probability $P$ is small, from Eq. (\ref{h2d}) we have:
\begin{eqnarray}\label{reg}
 &&\ln(P) \approx a - 2\nu \ln(S_0)\\
 &&a = 2\nu\ln(S_*)
\end{eqnarray}
Assuming a linear relationship between the CDS spread $Z$ and the default probability $P$, i.e., $Z\approx b~P$ (where $b$ is the aforesaid normalization factor, whose precise value will turn out not be relevant below), we have
\begin{eqnarray}\label{reg1}
 &&\ln(Z) \approx {\widetilde a} - 2\nu \ln(S_0)\\
 &&{\widetilde a} = a + \ln(b)
\end{eqnarray}
So, if we have a time series of CDS spreads $Z(t_s)$ (where $t_s$, $s = 1,\dots,M$, are the times in the time series, e.g., trading days), then we can run a linear regression of $\ln(Z(t_s))$ over the logs of the prices $\ln(S(t_s))$ (where $S(t_s)$ corresponds to $S_0$ in Eq. (\ref{reg1})) with the intercept. The regression coefficient of $\ln(S(t_s))$ is nothing but $-2\nu$. The catch is that in real life the drift and volatility are not constant, so running a regression over a long time period would make little sense, and if it is run over a short period (e.g., 1 month) with the view of capturing a short-horizon sentiment of the CDS market on the drift, then the intercept ${\widetilde a}$ (which is related to the ``threshold" price $S_*$) can, as is often the case, be expected to be unstable out-of-sample, which affects the forecasting power for the drift. {\em As{\'i} es la vida}. There is no free lunch.

{}However, instead of thinking about this model in terms of applying it to just a single stock, we can try a statistical approach by taking a large cross-section of stocks, extracting shorter-horizon expected returns for each stock as above, and then using these expected returns to construct, e.g., a dollar-neutral strategy, either via (constrained) optimization, or by ranking (e.g., buying the stocks in the top decile by $\nu$, and selling the stocks in the bottom decile). Such a cross-sectional approach, as in other strategies (see, e.g., \cite{151}), may reduce noise and improve the performance characteristics (e.g., the Sharpe ratio \cite{Sharpe}) of the trading strategy. So, this alpha can then be used in the context of statistical arbitrage, possibly in a mix of many other alphas (and with a caveat that CDS data may not be available for the entire universe of stocks traded by other alphas).

{}This write-up was inspired by a discussion at a dinner in December 2018 with Peter Carr and Igor Halperin (which discussion prompted me to make an indirect parallel of the topic of the healthy vs. distressed dynamics with \cite{Bubbles}), and a subsequent stimulating email correspondence with Igor Halperin, to whom I am grateful for discussing the ideas set forth in \cite{Halperin}.


\begin{thebibliography}{99}

\makeatletter
\def\@biblabel#1{}
\makeatother

\bibitem[Black and Scholes, 1973]{BS} Black, F. and Scholes, M. (1973)
The Pricing of Options and Corporate Liabilities.
{\em Journal of Political Economy} 81(3): 637-654.

\bibitem[Halperin and Dixon, 2018]{Halperin} Halperin, I. and Dixon, M. (2018)
``Quantum Equilibrium-Disequilibrium": Asset Price Dynamics, Symmetry Breaking, and Defaults as Dissipative Instantons.
{\em Working Paper}. Available online: \url{https://ssrn.com/abstract=3223243}.

\bibitem[Kakushadze, 2017]{Bubbles} Kakushadze, Z. (2017)
On Origins of Bubbles.
{\em Journal of Risk \& Control} 4(1): 1-30. Available online: \url{https://ssrn.com/abstract=2830773}.

\bibitem[Kakushadze and Serur, 2018]{151} Kakushadze, Z. and Serur, J.A. (2018)
{\em 151 Trading Strategies.} Cham, Switzerland: Palgrave Macmillan/Springer Nature.

\bibitem[Merton, 1973]{Merton1973} Merton, R. (1973)
Theory of Rational Option Pricing.
{\em Bell Journal of Economics and Management Science} 4(1): 141-183.

\bibitem[Merton, 1974]{Merton1974} Merton, R. (1974)
On the Pricing of Corporate Debt: the Risk Structure of Interest Rates.
{\em Journal of Finance} 29(2): 449-470.

\bibitem[Sharpe, 1994]{Sharpe} Sharpe, W.F. (1994)
The Sharpe Ratio.
{\em Journal of Portfolio Management} 21(1): 49-58.


\end{thebibliography}
\end{document}